\documentclass[prl,floatfix,preprint,,superscriptaddress,email,amsmath,amssymb]{revtex4-1}

\usepackage{graphicx}
\usepackage{amsmath,amssymb}
\usepackage{color}
\usepackage{psfrag}
\usepackage{ifsym}
\usepackage{epstopdf}
\usepackage{soul}
\usepackage{tikz}
\usepackage{tabularx}

\newcommand{\bra}[1] {\langle #1 |}
\newcommand{\ket}[1] {| #1 \rangle}

\begin{document}

\title{Coherent control of a solid-state quantum bit with few-photon pulses}

\author{V. Giesz$^*$}
\affiliation{CNRS-LPN Laboratoire de Photonique et de Nanostructures, Universit\'e Paris-Saclay, Route de Nozay, 91460 Marcoussis, France}
\author{N. Somaschi$^*$}
\affiliation{CNRS-LPN Laboratoire de Photonique et de Nanostructures, Universit\'e Paris-Saclay, Route de Nozay, 91460 Marcoussis, France}
\author{G. Hornecker$^*$}
\affiliation{Universit\'e Grenoble Alpes, F-38000 Grenoble, France}
\affiliation{CNRS, Institut N\'eel, “Nanophysique et Semiconducteurs” Group, F-38000 Grenoble, France}

\author{T. Grange}
\affiliation{Universit\'e Grenoble Alpes, F-38000 Grenoble, France}
\affiliation{CNRS, Institut N\'eel, “Nanophysique et Semiconducteurs” Group, F-38000 Grenoble, France}

\author{B. Reznychenko}
\affiliation{Universit\'e Grenoble Alpes, F-38000 Grenoble, France}
\affiliation{CNRS, Institut N\'eel, “Nanophysique et Semiconducteurs” Group, F-38000 Grenoble, France}

\author{ L. De Santis$^*$}
\affiliation{CNRS-LPN Laboratoire de Photonique et de Nanostructures, Universit\'e Paris-Saclay, Route de Nozay, 91460 Marcoussis, France}
\affiliation{Universit\'e Paris-Sud,  Universit\'e Paris-Saclay, F-91405 Orsay, France}

\author{J. Demory}
\affiliation{CNRS-LPN Laboratoire de Photonique et de Nanostructures, Universit\'e Paris-Saclay, Route de Nozay, 91460 Marcoussis, France}
\author{C. Gomez}
\affiliation{CNRS-LPN Laboratoire de Photonique et de Nanostructures, Universit\'e Paris-Saclay, Route de Nozay, 91460 Marcoussis, France}
\author{I. Sagnes}
\affiliation{CNRS-LPN Laboratoire de Photonique et de Nanostructures, Universit\'e Paris-Saclay, Route de Nozay, 91460 Marcoussis, France}

\author{A. Lemaitre}
\affiliation{CNRS-LPN Laboratoire de Photonique et de Nanostructures, Universit\'e Paris-Saclay, Route de Nozay, 91460 Marcoussis, France}

\author{O. Krebs }
\affiliation{CNRS-LPN Laboratoire de Photonique et de Nanostructures, Universit\'e Paris-Saclay, Route de Nozay, 91460 Marcoussis, France}

\author{N. D. Lanzillotti-Kimura}
\affiliation{CNRS-LPN Laboratoire de Photonique et de Nanostructures, Universit\'e Paris-Saclay, Route de Nozay, 91460 Marcoussis, France}
\author{L. Lanco}
\affiliation{CNRS-LPN Laboratoire de Photonique et de Nanostructures, Universit\'e Paris-Saclay, Route de Nozay, 91460 Marcoussis, France}
\affiliation{D\'epartement de Physique, Universit\'e Paris Diderot, 4 rue Elsa Morante, 75013 Paris, France}

\author{A. Auffeves}
\affiliation{Universit\'e Grenoble Alpes, F-38000 Grenoble, France}
\affiliation{CNRS, Institut N\'eel, “Nanophysique et Semiconducteurs” Group, F-38000 Grenoble, France}

\author{P. Senellart}
\affiliation{CNRS-LPN Laboratoire de Photonique et de Nanostructures, Universit\'e Paris-Saclay, Route de Nozay, 91460 Marcoussis, France}
\affiliation{D\'epartement de Physique, Ecole Polytechnique, Universit\'e Paris-Saclay, F-91128 Palaiseau, France}

$*$ equally contributing authors
\begin{abstract}
{\bf Single photons are the natural link between the nodes of a quantum network: they coherently propagate \cite{quantuminternet} and interact with many types of quantum bits including natural and artificial atoms \cite{ionphoton,atomphoton,NVphoton,QDphoton}. Ideally, one atom should deterministically control the state of a photon and vice-versa.  The interaction between free space photons and an atom is however intrinsically weak and many efforts have been dedicated to develop an efficient interface \cite{BookAna}. 
Recently, it was shown that the propagation of light  can be controlled by an atomic resonance coupled to a cavity or a single mode waveguide \cite{atacgate, waksfewphoton,jelenafew,loo,lukin2013,gaterempe,barak}. 
Here we demonstrate that the state of a single artificial atom in a cavity can be efficiently controlled by a few-photon pulse. We study a quantum dot optimally coupled to an electrically-controlled cavity device \cite{dousse}, acting as a near optimal one-dimensional atom \cite{1Datom}. By monitoring the exciton population  through resonant fluorescence, we demonstrate Rabi oscillations with a $\pi$-pulse of only 3.8 photons on average. The probability to flip the exciton quantum bit with a single photon Fock state is calculated to reach 55\% in the same device.
}
\end{abstract}

\maketitle

Photonic channels are close to ideal transmitters of the quantum information and the light-matter interaction provides a natural framework to interface photonic channels and quantum nodes. Photons interact with a large variety of quantum systems, from atomic ensembles \cite{memory} to single atoms \cite{atommemory} or ions \cite{ionphoton}, as well as all kinds of solid-state artificial atoms such as semiconductor quantum dots \cite{atacgate,waksfewphoton,jelenafew,loo} or defects in diamonds \cite{NVcoherentcontrol}. The scalability of a quantum network requires highly efficient quantum interfaces between flying and stationary quantum bits.  However, mapping a photonic channel to a single (artificial) atom in free space is naturally inefficient: using strong light focusing and temporally shaped laser pulses, 50 photons  are still typically needed to  control an atomic state \cite{50photons}. An ideal atom-photon interface requires that a single atom interacts with only a single and well-defined mode of the optical field.  Such ideal “one-dimensional (1D) atom” situation can be obtained by controlling the spontaneous emission of the atom using photonics structures like single sided leaky cavities \cite{gazzano2013,lukin2013}, single mode photonic waveguides \cite{claudon2010,lodahl1D} or  fibers evanescently coupled to directional cavity modes \cite{barak}: the spontaneous emission is accelerated into a well defined mode of the electromagnetic field \cite{gazzano2013,lukin2013,barak} or inhibited in all other modes \cite{claudon2010,lodahl1D}. Spectacular results have been reported in this area in the last few years. Ultrabright single photon sources have been obtained by placing semiconductor quantum dots (QDs) in cavities \cite{gazzano2013} or single mode nanowires \cite{claudon2010,lodahl1D}. Single natural or artificial atoms in macroscopic cavities or photonics nanostructures have been shown to control the reflectivity or transmission of attenuated coherent pulses, with first demonstrations of optical gates operating at the single photon level \cite{atacgate, waksfewphoton, jelenafew,loo,lukin2013, gaterempe,barak}.  Elementary quantum networks have been demonstrated with atoms in macroscopic cavities, where the photon emitted by one atom is absorbed by a second one \cite{networkrempe}.  Here, we report  on the  coherent control of a near optimal 1D artificial atom with few-photon pulses.

We study the coherent control of a  QD quantum bit coupled to a microcavity. Two types of quantum bits are explored with single QDs: excitons \cite{excitonqbit} or spin quantum bits \cite{spinqbit}. Here we study an exciton transition optimally coupled to a micropillar cavity.  Light is confined vertically in a $\lambda$-cavity surrounded by two GaAs/Al$_{0.9}$Ga$_{0.1}$As Bragg-reflectors and laterally through the etching of a connected pillar structure as introduced in ref. \cite{nowak}. Figure 1a presents a sketch of the device: a single quantum dot is inserted in a circular micropillar cavity with diameter 2.9 $\mu$m, connected through one-dimensional 1.4 $\mu m$ wide ridges to a large frame where an electrical contact is defined. Combined with a p-i-n doping of the layers, this geometry allows applying a bias to the structure, a critical tool to finely tune the QD resonance to the cavity mode energy as well as to stabilize the charge environment of the QD \cite{nearoptimal}. 
The cavity is fabricated using a cryogenic in-situ lithography  to center each cavity device on a single QD with 50 nm spatial accuracy \cite{dousse}. Evidence for this deterministic positioning is provided by the emission map of the device shown in figure 1b: a very bright signal originating from the QD appears at the center of the cavity, signature of increased collection efficiency.  

 \begin{figure}[h!]
		\begin{center}
		\includegraphics [width= 15cm]{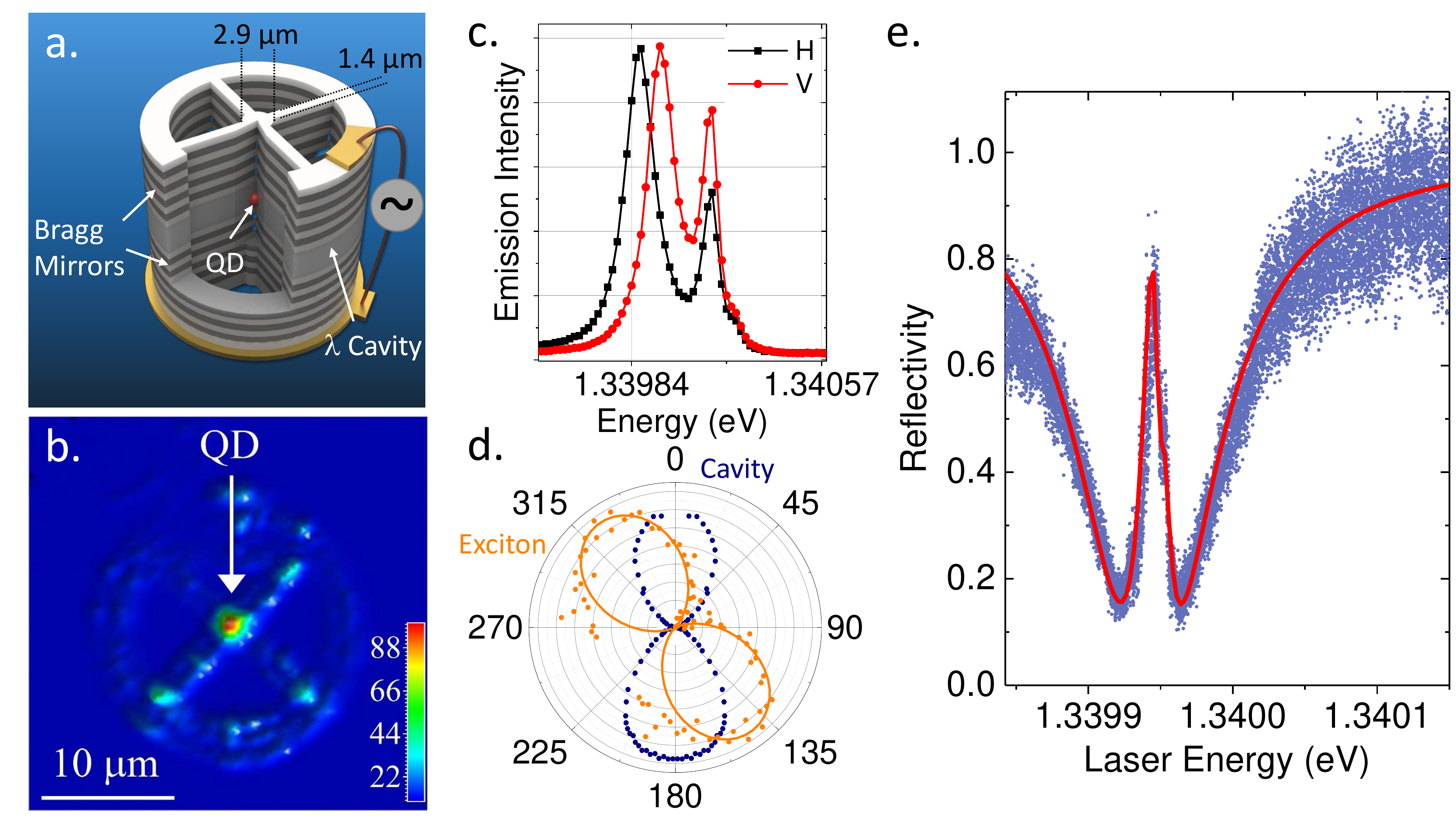}
		\end{center}
\caption{{\bf A quantum dot in a cavity as a one dimensional atom system: } a: Sketch of the device: a single QD is located at the center of a pillar cavity connected through 1.4 $\mu$m wide ridges to a circular frame where the electrical contact is defined. b: Emission map of the device obtained by  scanning the sample below the excitation laser spot. c: Emission spectra measured at 0 V bias under non-resonant excitation at 850 nm for two linear polarizations for the detection, labeled H and V. d : Polar plot of the QD exciton energy (orange) and of the cavity mode energy (blue)  deduced from polarization resolved emission measurements (see c). e: Reflectivity measured for an excitation power around 50 pW by scanning a V-polarized monomode continuous wave laser across the cavity resonance at the bias of resonance with the QD. Blue: measured reflectivity. Red: theoretical fit (see supplementary).  
}\label{fig1}
	\end{figure}

The sample is placed in a closed-cycle cryostat at 4K and  a confocal geometry setup is used to excite the QD and collect the signal through a microscope objective (NA 0.75). Figure 1.c presents emission spectra of the device at zero bias under non-resonant excitation at 850 nm, collected  in two linear polarizations labeled H and V. Each spectrum presents two emission lines, one corresponds to the cavity mode (CM) the other to the QD exciton line. Both line energies depend on the detected polarization.  The fundamental mode of the pillar around 1.3399eV  decomposes into two linearly-polarized modes (H and V) split by 70 $\mu$eV with a  linewidth of $\kappa=120 \ \mu eV$  (quality factors around 11000) whereas the exciton fine structure splitting  amounts to $\Delta_{FSS} =15\ \mu eV$. As shown by the polar plot in figure 1d, the QD axes labeled X and Y are roughly aligned to the crystal axes at 40$^\circ$ from the cavity ones. This configuration allows us to monitor the exciton population using resonant fluorescence as explained below. 

To characterize the QD-cavity coupling, reflectivity measurements are performed scanning a V-polarized continuous wave laser across the cavity mode resonance. Figure 1e, obtained for an applied bias corresponding to the cavity-exciton resonance, evidences a strong coherent response of the QD at the center of the cavity dip \cite{loo}. Using a V-polarized excitation and collection, only the V cavity mode is visible while both exciton dipoles polarized along X and Y are excited, giving rise to a slightly asymmetric QD-induced reflectivity peak. A theoretical adjustment shown in red, which considers the two exciton states and the  two cavity modes (see supplementary), evidences a cavity-QD coupling strength $g=21 \mu eV$ and a  radiatively limited dephasing rate $\gamma=0.3 \mu eV$ for the exciton. The derived cooperativity is as large as $C=\frac{g^2}{\kappa \gamma}=13$: it is obtained owing to a total suppression of charge noise in these gated  structures, an observation consistent with the emission of fully indistinguishable photons from similar devices \cite{nearoptimal}.  At high excitation power, the QD response saturates and only the reflectivity dip of the bare cavity is visible (see supplementary). The minimum reflectivity then provides the fraction of photons escaping the cavity mode through the top mirror, i.e. the  out-coupling efficiency, $\eta_{out}=0.7 \pm 0.05$. Finally, we measure  the mode profile of the incoming beam and  that of the pillar fundamental mode, and deduce an excellent mode matching corresponding to an input coupling efficiency exceeding $\eta_{in}>0.95$. These measurements show that the present device is close to the textbook situation of the 1D atom: every photon sent on the device enters the cavity with probability $\eta_{in}>0.95$ and interacts with the exciton. The exciton emits a photon back into the cavity mode with a probability $\frac{2C}{2C+1}=0.96$ and this photon escapes through the top mirror with a probability  $\eta_{out}=0.7 \pm 0.05$. As discussed below,  this 1D atom configuration allows a coherent control of the exciton quantum bit with few photon pulses.

The quantum bit state is probed by resonantly exciting the device with shaped pulses from a Titanium-Sapphire laser delivering 3 ps pulses at a 82 MHz repetition rate. The pulse  duration is adjusted between 10 ps and 90 ps. Using a polarizing beam splitter (PBS) and a half waveplate,  we excite the device with linearly polarized light along V and separate the H and V polarized device optical response  (figure 2a). The exciton population is monitored by measuring the exciton fluorescence in the orthogonal H polarization.

 \begin{figure}[h!]
		\begin{center}
		\includegraphics [width= 15cm]{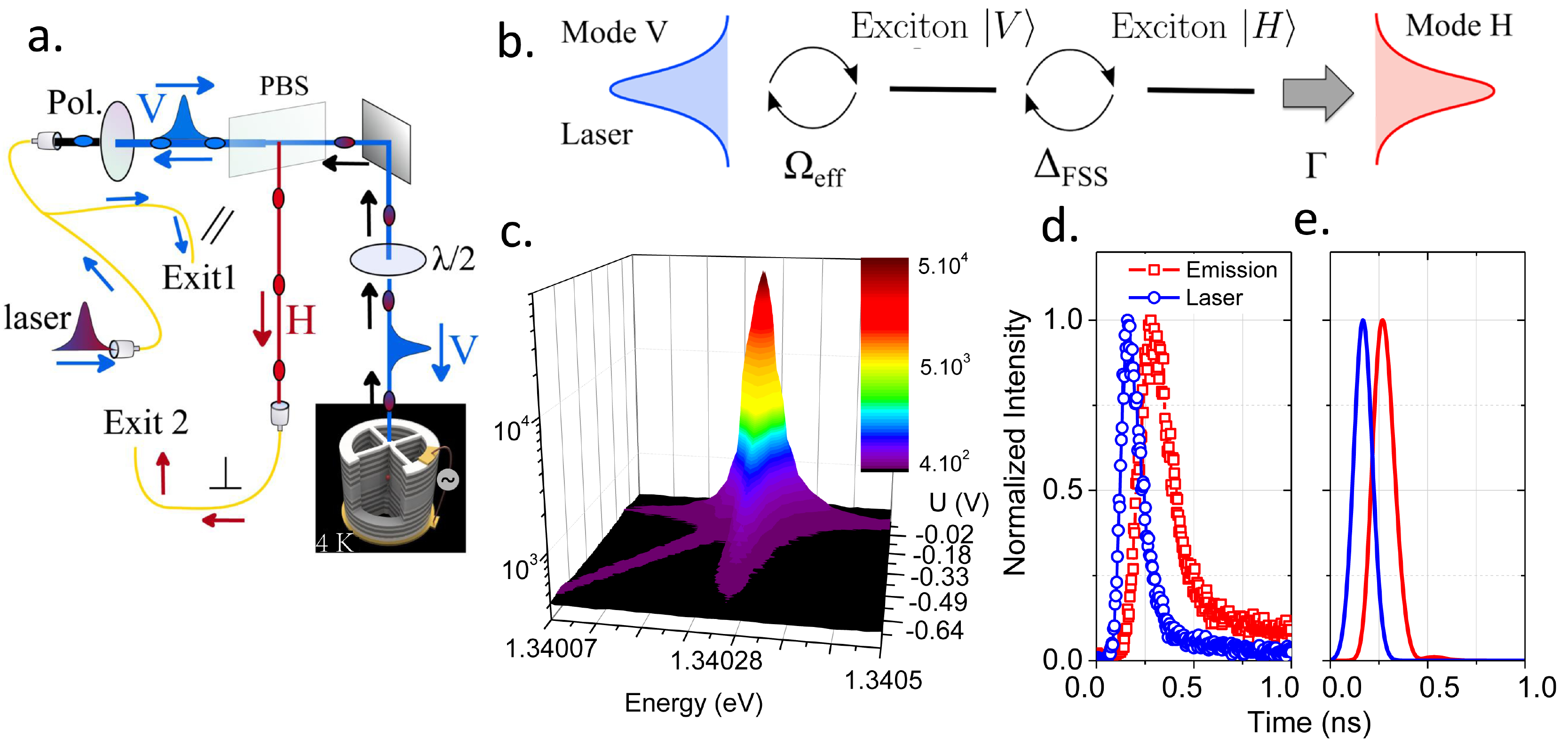}
		\end{center}
\caption{{\bf Resonant fluorescence measurement used to monitor the exciton population: } a: Sketch of the experimental setup: the laser polarization is controlled by a polarizing beam splitter and a half waveplate. The reflectivity is measured by collecting the signal at exit 1 while the resonant fluorescence  is measured at exit 2 in crossed polarization. b : Schematic representation of the theoretical model used to describe the experiment. See text for details. c : Emission intensity (log. scale) measured in crossed polarization as a function of energy and applied bias. The laser is resonant to the V mode energy.   d, e: Time dependence of the 56 ps excitation pulse (blue) and the  collected emission in H polarization (red).  d: experiment e: calculations.  }\label{fig2}
	\end{figure}

Figure 2c presents emission spectra collected in H polarization when continuously varying the bias applied on the device. The laser energy is set to the energy of the V polarized cavity mode.  By increasing the applied voltage, the QD excitonic transition is tuned in resonance with the laser energy where a strong fluorescence signal is observed. Away from resonance, a faint emission at the exciton energy is still observed due to phonon assisted processes. The mechanism leading to an exciton emission in crossed  polarization can be seen as follows: the laser drives the V polarized exciton state  $\ket{\psi(t=0)}=\ket{V}$ which correspond to a linear superposition of X and Y excitons state $\ket{V}=\frac{\ket{X}+\ket{Y} }{\sqrt{2}}$. This superposition temporally evolves  as $\ket{\psi(t)}=\frac{ \ket{X}+e^{-i \Delta_{FSS} t/\hbar}\ket{Y}}{\sqrt{2}}$, eventually leading to an emission along  H polarization. To theoretically account for the experiment,  a model schematically described in fig 2b and presented in details in the supplementary materials is developed. The QD is modeled as a V-type 3-level system (i.e. the ground state $\ket{g}$ and the two polarized excitons $\ket{X}$ and $\ket{Y}$).
Its interaction with  the V and H polarized cavity modes is treated within the rotating-wave approximation.
When the V-mode cavity is driven by a coherent field, the density matrix of the system consisting of the V-mode coupled to the QD is calculated using a Lindblad master equation (see supplementals for details).
Figure 2d-e presents the measured and calculated time dependences of the 56 ps excitation pulse  (blue) and the  H-polarized exciton emission (red). The emission is temporally delayed from the excitation pulse in agreement with theoretical predictions: the smaller the FSS, the longer the temporal evolution toward the orthogonal exciton state.

 \begin{figure}[h!]
		\begin{center}
		\includegraphics [width= 15cm]{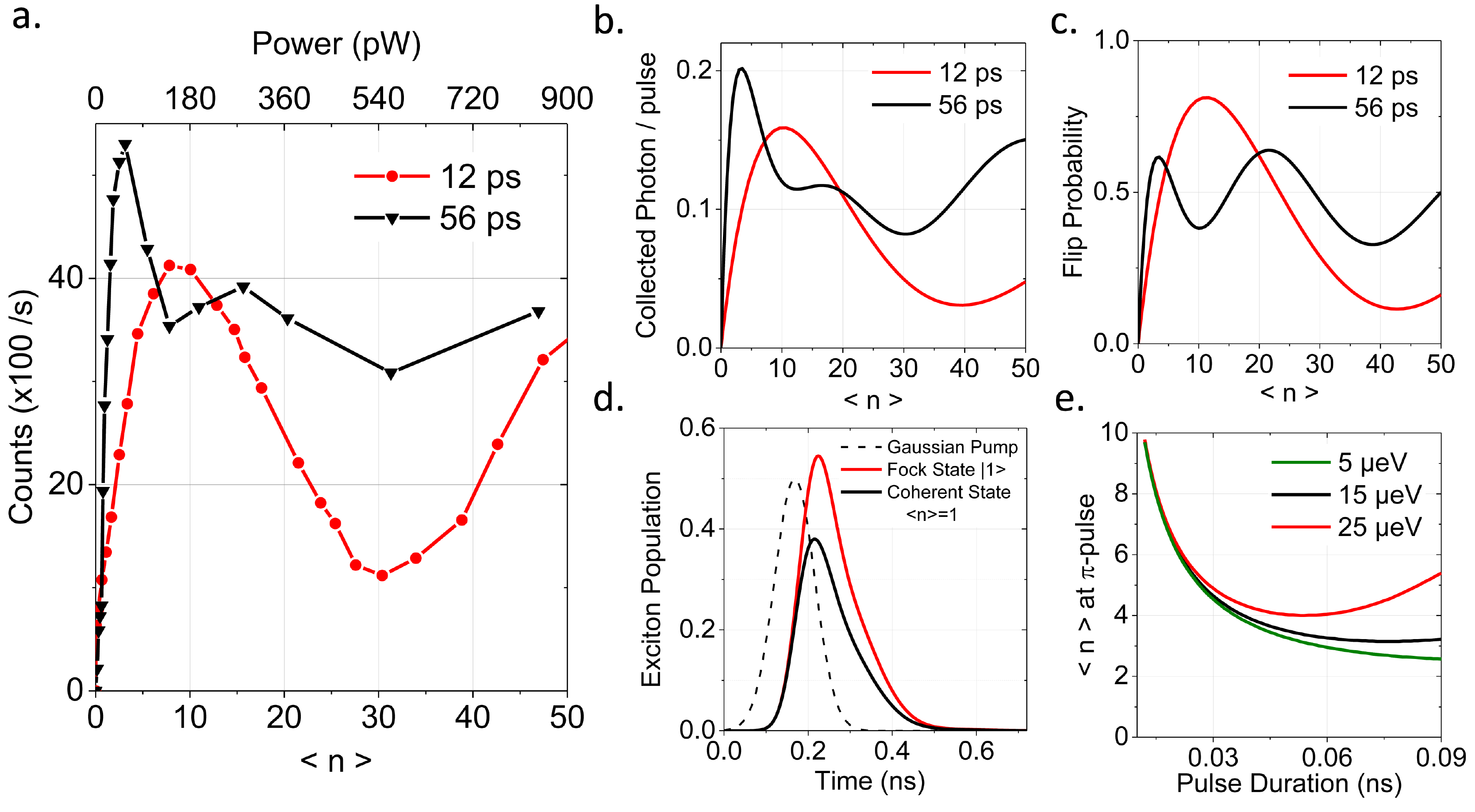}
		\end{center}
\caption{{\bf Coherent control with few photon pulses: }
a: H-polarized emission  intensity as a function of the excitation power (top axis) and the average photon number $\langle n \rangle$ (bottom axis) for two pulse durations (12 ps and 56 ps). b: Calculated H-polarized emission as a function of $\langle n \rangle$ for two pulse durations. c: Calculated probability to find the QD in its excited state after the pulse as a function of $\langle n \rangle$  for 12 ps and 56 ps pulses. d: Calculated probability to find the QD in the exciton state as a function of time for a 56 ps excitation pulse (dashed line). Black : case of a coherent pulse with $\langle n \rangle=1$. Red : case of a single photon Fock state. e: Mean photon number $\langle n \rangle$ needed to induce a $\pi$-pulse as a function of the pulse duration for various values of $\Delta_{FSS}$. }\label{fig3}
	\end{figure}

We now study the coherent control of the exciton by monitoring the time integrated H-polarized emission intensity.  The exciton population oscillates during the pulse, driven by the V polarized excitation. The H-polarized emission  mostly takes place after the pulse as shown in figure 2d. When increasing the average power for a given pulse duration, the probability for the QD to be in the exciton state at the end of the pulse oscillates \cite{rabiQD,valia},  resulting in oscillations in the H-polarized emission with the excitation power. Figure 3a presents the time integrated H-polarized  emission intensity  as a function of the average power sent on the device (top scale). Clear Rabi oscillations are evidenced  for  both 12 and 56 ps pulses.   The bottom scale of figure 3a shows the mean photon number $\langle n \rangle$ per pulse sent on the device, derived from the average excitation power $P$, the laser repetition rate $1/\Gamma_{rep}=12\ ns$ and the photon energy $E=1.34 \ eV$: $\langle n \rangle=\frac{P}{\Gamma_{rep} E}$. Only 3.8  (8.6) photons are sent on average at $\pi$-pulses for a pulse duration of 56 (12) ps. This extremely low photon number is obtained owing to the excellent mode matching between the incident gaussian beam and the fundamental mode of the micropillar,  in contrast with photonic crystal cavities where the mode profile is difficult to control \cite{atacgate,waksfewphoton,jelenafew}. Experimental observations are theoretically reproduced in Figure 3b presenting the calculated H-polarized exciton emission  as a function of $\langle n \rangle$. The model includes only the parameters extracted from figure 1e and assumes a perfect input coupling efficiency ($\eta_{in}=1$). An excellent agreement with the experimental observations is obtained. Though these Rabi oscillations demonstrate the few-photon coherent control of the exciton, it does not provide a direct measure of the probability to flip the exciton state. This flip probability can be  derived from the theoretical model by calculating the  probability of finding the QD in its excited state (both H or V dipole) as shown in figure 3c for the two pulse durations. At $\pi-$pulse, sending 8.6 (3.8) photons is sufficient to flip the exciton qubit with a probability of 81\%  (62\%) depending on the pulse length. The  flip probability is sligthly reduced for the longer pulse: Rabi oscillations driven by the V excitation are damped by the spontaneous emission of the exciton in the V mode and by the time evolution toward the $\ket{H}$ exciton state. With a Purcell factor of 13, corresponding to an emission decay time around 100 ps, spontaneous emission damping hardly plays a role for the 12 ps pulse, while oscillations are  damped for a 56 ps pulse. On the other hand, more photons are needed with the 12 ps pulse, because of the lower spectral overlap between the excitation pulse and the exciton state.

Such  low photon number coherent control establishes a new state of the art for solid-state optical devices. Until now, such efficient interface had only been reached with a single atom trapped in a cavity \cite{gaterempe} or close to fiber-coupled microresonator \cite{barak}. It is demonstrated here for a  micron-size semiconductor device produced with a fully deterministic and highly  reproducible technology.
To flip the exciton quantum bit with even lower photon numbers, two parameters should be considered. First, the current experiment uses attenuated coherent pulses: for a coherent pulse with $\langle n \rangle=1$, the probability that the pulse actually contains no photons is 1/e=36\%. Figure 3d presents the population of the QD excited state as a function of time comparing the case of an incident coherent state with $\langle n \rangle=1$ and the case of a single photon Fock state. To account for an incoming single photon wavepacket, the generalized master equation of Ref.~\cite{1photon} is used (see supplementary). A 44\% increase of the probability to flip the quantum bit is expected: sending a single photon on the same  device would ensure a probability to flip the exciton quantum bit with $55 \%$ probability. 
The other parameter that increases the number of photons at $\pi$-pulse is specific to the present experiment. Since the exciton state presents a sizeable FSS, around $\Delta_{FSS}=15 \mu$eV, the V exciton dipole rotates toward the H exciton dipole where efficient spontaneous emission into the H cavity mode takes place. This mechanism contributes to damp the exciton population during the excitation, leading to a higher  $\langle n \rangle$ at $\pi$-pulse. Figure 3f presents the average number of photon at $\pi$-pulse for three values of $\Delta_{FSS}$, as a function of the pulse duration. For each value of the $\Delta_{FSS}$, the mean photon number $\langle n \rangle$ at $\pi$-pulse presents a minimum value. The smaller $\Delta_{FSS}$, the lower the minimal photon number $\langle n \rangle$ is and the longer the optimal pulse. 

We have reported on a highly efficient solid state interface between a flying  and a stationary quantum bit. This has been obtained by deterministically inserting a QD in an electrically controlled pillar cavity that ensures both an excellent mode matching with incident Gaussian beams and a state of the art cooperativity of $C=13$. The device presented here has been designed to operate with a neutral exciton. Similar cavities, coupled to a positively charged exciton, bring exciting perspectives to demonstrate new quantum functionalities. Since the charged exciton optical resonance can be used to initialize the hole spin \cite{holespin}, manipulating a single spin with only a single photon is foreseeable using the present technology \cite{loicini}. Such efficient spin-photon interface would open the way to deterministic quantum gates between delayed photons as well as cluster state generation \cite{rarity}. 

 
\vspace{0.5 cm}

{\bf Acknowledgments:} This work was partially supported by the ERC Starting Grant No.
277885 QD-CQED, the French Agence Nationale pour la Recherche (grant ANR QDOM and SPIQE)
the French RENATECH network, the Labex NanoSaclay, and the EU FP7 Grant No.
618072 (WASPS), N.D.L.K. was supported by the FP7 Marie Curie Fellowship OMSiQuD. 

\vspace{0.5 cm}

{\bf Author contributions:}
Optical measurements were conducted primarily by V.G. N.S. and
L.d.S., with help from J.D., L.L., O.K. and P.S. The electrically controlled samples were
fabricated by N.S.  The sample was grown by C.G. and A.L., and the
etching performed by I.S. Theory was developed by G.H., T.G., B. R. and A.A. The project was conducted by P.S. on the experimental side  and by A.A. on the theory side. All authors discussed the results and participated to manuscript
preparation.

{\bf Author information:} Correspondence and requests for materials should be addressed
to Pascale.Senellart@lpn.cnrs.fr and Alexia.Auffeves@neel.cnrs.fr.


\pagebreak
\widetext
\begin{center}
	\textbf{\large Supplementary Material of\\ Coherent control of a solid-state quantum bit with few-photon pulses}
\end{center}
\setcounter{equation}{0}
\setcounter{figure}{0}
\setcounter{table}{0}
\setcounter{page}{1}
\makeatletter
\renewcommand{\theequation}{S\arabic{equation}}
\renewcommand{\thefigure}{S\arabic{figure}}
\renewcommand{\bibnumfmt}[1]{[S#1]}
\renewcommand{\citenumfont}[1]{S#1}

\section{Measurements}
\subsection{Sample}
The microcavity sample was grown by molecular beam epitaxy. A $\lambda$-GaAs cavity is surrounded by a  bottom and a top mirror made of 30 and 20 pairs of  $Al_{0.95}Ga_{0.05}As/GaAs$. The bottom mirror is gradually  $n$-doped while the top one is $p$-doped. After the in-situ optical lithography defining the cavities centered on the QDs, the sample is etched and standard $p$ contacts are defined on a large frame ($\approx 100 \times 100\mu m^2$) connected to the circular frame around the micropillar (see Fig. 1a and Ref. \cite{nowak}). A standard $n$ contact is defined on the back of the sample.

\subsection{Experimental setups}
The experimental setup is based on a confocal geometry where the same microscope objective (NA $0.75$) serves simultaneously for quantum dot (QD) excitation and photoluminescence (PL) emission collection. 
The excitation is provided by a tunable Ti-Sapph laser, providing  3 ps pulses at a $82$ MHz repetition rate.  The sample is kept at  $4$ Kelvin in a close-cycle cryostat with an exchange gas. 

The laser pulse duration is  adjusted by sending 3 ps laser pulses through a spectrometer and an etalon filter with 10 pm bandwidth. By adjusting the spectrometer slit size and the etalon angle, the pulse width can be continuously tuned between 10 ps and 100 ps. 

\subsection{Reflectivity measurement}
	\begin{figure}[htp]
		\begin{center}
			\includegraphics [width= 10cm]{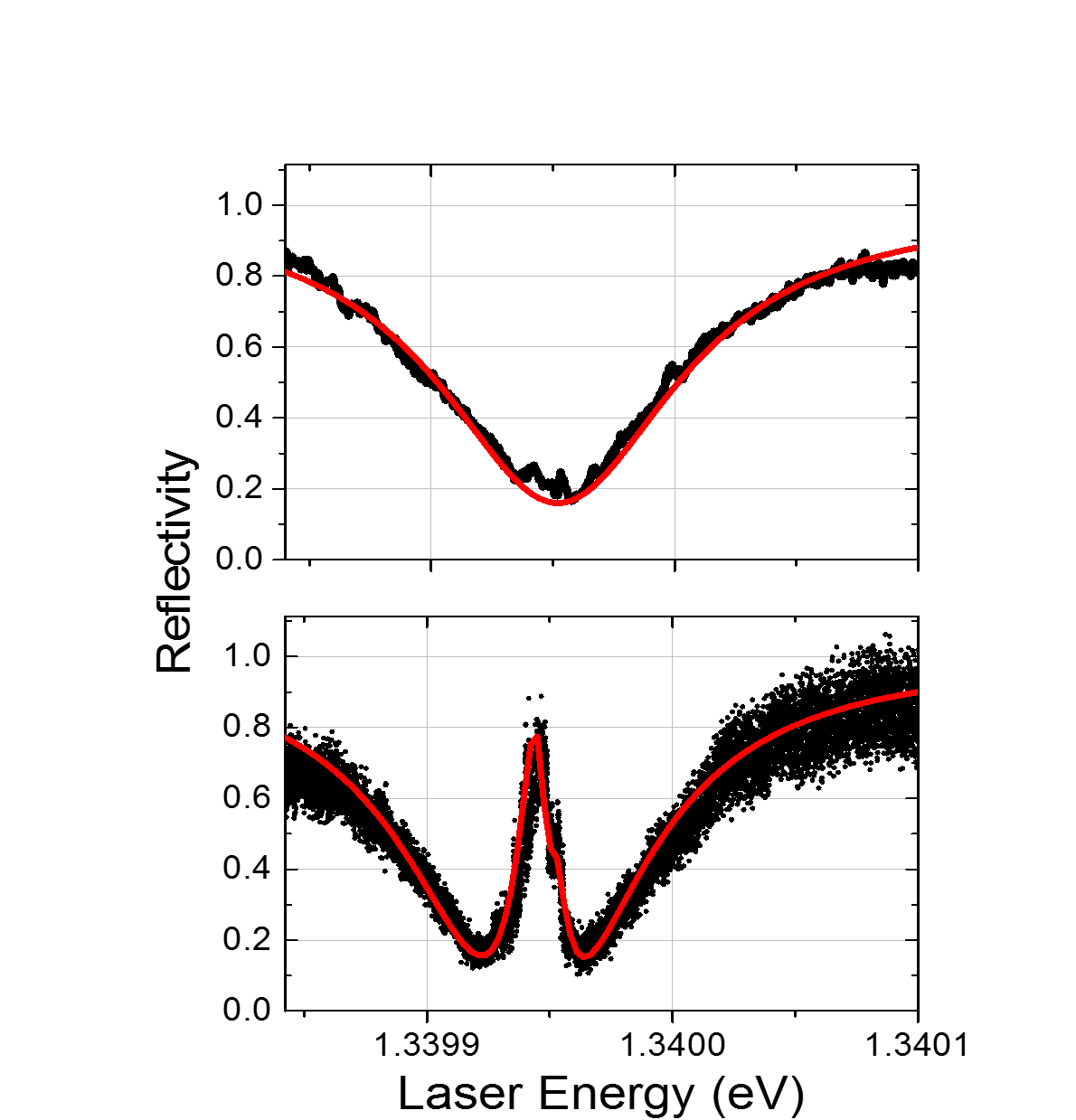}
		\end{center}
		\caption{\textbf{Reflectivity.}  Reflectivity spectra and relative fitting measured for a bias where the QD exciton transition is detuned from the cavity mode (top) or in resonance with the cavity mode (bottom). }\label{fig:SMQD4}
	\end{figure}

To measure the reflectivity of the micropillar, a CW tunable laser focused on the pillar is reflected and sent to a detector (Exit 1 in Fig 2a). The intensity of the reflected light is collected and measured as a function of the laser energy. When the exciton transition is not in resonance with the mode of the cavity, a reflectivity dip is observed corresponding to the bare cavity resonance (Fig. S1, top). The width of the cavity mode gives an accurate value of the cavity damping $\kappa = 120 \mu eV$ corresponding to a Q-factor equal to 11000. From the value of the reflectivity at the cavity energy $R_{min}$, we can extract the coupling of the pillar through the top mirror $\kappa_{top}/\kappa$ : 
$R_{min}=(1-2 \kappa_{top}/\kappa)^2$.
We extract two possible values for the output coupling $\kappa_{top}/\kappa = 0.7 \pm 0.05$ or $\kappa_{top}/\kappa = 0.3 \pm 0.05$. Considering that the mirrors of the cavity are highly asymmetric (with 30/20 pairs) and that the same device acts as a bright single photon source \cite{nearoptimal}, we deduce that $\kappa_{top}/\kappa = 0.7 \pm 0.05. $

By changing the electrical applied bias, the exciton transition is tuned in resonance with the cavity mode. The exciton transition appears as the narrow peak in the middle of the cavity resonance (Fig. S1, bottom). The height of the peak depends on the coupling strength between the QD and the cavity $g$ and on the coherence rate $\gamma=\gamma_{sp}/2 + \gamma^*$, where $\gamma_{sp}$ is the spontaneous emission rate in the other modes and $\gamma^*$ the pure dephasing rate \cite{loo}. By fitting the experimental data, we extract a coupling strength equal to $g=21 \pm 2 \mu eV$ and a radiatively limited coherence rate equal to $\gamma= 0.30 \pm 0.10 \mu eV$. The device cooperativity $C=g^2/\kappa \gamma$ reaches the very high value of $C = 13$. 

\section{Theory}

\begin{figure}[h!]
	\begin{center}
		\begin{tikzpicture}
		\node[thick,align=center] at (-0.2,2) {(a)};
		\draw[-,blue,very thick] (0.3,0) -- node[above] {$\ket{G}$} (2.7,0);
		\draw[-,blue,very thick]  (1.8,1.7) -- node[above] {$\ket{Y}$} (3,1.7);
		\draw[-,blue,very thick] (0,1.3) -- node[above] {$\ket{X}$} (1.2,1.3);
		\draw[|<->|, red] (0.7,0.01) --node[right, black]{$\omega_X$} (0.7,1.29); 
		\draw[|<->|, red] (2.5,0.01) --node[right, black]{$\omega_Y$} (2.5,1.68);
		\draw[|<->|,red] (1.3,1.29) -- node[right, black] {$\Delta_{FSS}$} (1.3,1.7);
		\node[thick,align=center] at (3.8,2) {(b)};
		\draw[-,blue,very thick] (4.3,0) -- node[above] {$\ket{G}$} (6.7,0);
		\draw[-,blue,very thick]  (5.8,1.5) -- node[above] {$\ket{V}$} (7,1.5);
		\draw[-,blue,very thick] (4,1.5) -- node[above] {$\ket{H}$} (5.2,1.5);
		\draw[|<->|, red] (4.5,0.01) --node[right, black]{$\omega_H$} (4.5,1.48) ;
		\draw[|<->|, red](6.5,0.01) -- node[right, black]{$\omega_V$} (6.5,1.48);
		\draw[<->, black, very thick](5.75,1.5)-- node[below, black]{$\Delta_{FSS}/2$}  (5.25,1.5);
		\end{tikzpicture}
	\end{center}
	\caption{Schematic of the quantum dot states considered in the model: (a) in the basis of the QD exciton eigenstates ($|X\rangle$,$|Y\rangle$); (b) in the ($|V\rangle$,$|H\rangle$) basis corresponding to the orientation of the cavity modes. For $\theta=\pi/4$ the coupling between $|V\rangle$ and $|H\rangle$ is $\Delta_{FSS}/2$. 
	} \label{QD}
\end{figure}
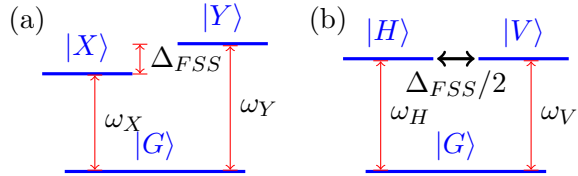

The quantum dot (QD) is modeled as a three-level system with a ground state $\ket{G}$ and two excited states $|X\rangle$ and $|Y\rangle$ corresponding to two linearly polarized excitonic states (See Figure~\ref{QD}). 
The respective energies of the excitons are denoted $\omega_X$ and $\omega_Y=\omega_X+\Delta_{FSS}$ where $\Delta_{FSS}$ is the fine structure splitting. The QD interacts with the two orthogonally polarized cavity modes $H$ and $V$ with respective lowering operators denoted $a_H$ and $a_V$, of respective frequencies $\omega_H$ and $\omega_V$.
The exciton eigenstates are transformed into the cavity-polarization basis through a rotation by an angle $\theta$: $\ket{V} = \cos(\theta) \ket{X} + \sin(\theta) \ket{Y}$ and  $\ket{H} = -\sin(\theta) \ket{X} + \cos(\theta) \ket{V}$. 
In this basis, the Hamiltonian of the QD-cavity system reads
\begin{align} \label{H}
	\hat{H}_\text{s}=\hat{H}_\text{QD}+\hat{H}_\text{c}+\hat{H}_\text{i} ,
\end{align}
where $\hat{H}_{QD}=\hbar (\delta_V^{at} \sigma_V^\dag\sigma_V+\delta_H^{at} \sigma_H^\dag\sigma_H)+ \Delta_{FSS} \cos(\theta) \sin(\theta) (\sigma_H^\dag\sigma_V+\sigma_V^\dag\sigma_H)$  is the free Hamiltonian of the QD, and $\delta_V^{at}=\delta_X \cos^2(\theta)+\delta_Y \sin^2(\theta)$ and $\delta_H^{at}=\delta_X \sin^2(\theta)+\delta_Y\cos^2(\theta)$ where $\delta_X$ and $\delta_Y$ are detunings from the pump frequency, which we shall take as the reference. We have introduced $\sigma_V = \ket{G}\bra{V}$ and $\sigma_H=\ket{G}\bra{H}$.  $\hat{H_c}=\hbar( \delta_V a_V^\dag a_V+\delta_H a_H^\dag a_H )$ is the free Hamiltonian of the cavity modes, 
and $ \hat{H_i}=\hbar g(a_V \sigma_V^\dag+a_H \sigma_H^\dag+a_V^\dag \sigma_V+a_H^\dag \sigma_H)$ is the QD-cavity interacting term within the rotating-wave approximation.\\

Each exciton can relax towards the vacuum with a spontaneous emission rate $\gamma$. On the other hand, each cavity mode is coupled to a 1D continuum of modes and to a reservoir of lossy modes of same polarization, giving rise to a decay rate denoted $\kappa$ and $\kappa_\text{loss}$ respectively.

\subsection*{Excitation by a coherent field}

When the cavity is pumped with a coherent $V$-polarized field through the 1D channel, it gives rise to an additional term in the {Hamiltonian}  $\hat{H}_p=\hbar (\Omega^*(t) a_V + \Omega(t) a_V^\dag)$ (see e.g. \cite{cohen-tannoudji}). The time-dependent Rabi frequency reads $\Omega(t)=\sqrt{n\kappa} \xi(t)$,
where  $n$ is the mean number of incident photons in the pulse, and the pulse $\xi(t)$ is a normalized Gaussian function checking \begin{equation}
	\xi(t)=\left(\dfrac{8 \ln(2)}{\pi \tau^{2}}\right)^{1/4}\exp(-{t^2}/{\tau^2}4\ln(2)).
\end{equation}
Introducing $\rho$ the density matrix of the coupled QD-cavity system, the Liouville equation ruling the dynamics writes:
\begin{equation} \label{L}
	\dot{\rho}=\mathcal{L}[\rho]=-\frac{i}{\hbar}\left[\hat{H}_s+\hat{H}_p,\rho\right]+ D_{\gamma,\sigma_H}[\rho] + D_{\gamma,\sigma_V}[\rho] + D_{\kappa_\text{tot},a_H}[\rho] + D_{\kappa_\text{tot},a_V}[\rho] ,
\end{equation}
where $D_{\alpha,X}[\rho] = \alpha \left( X \rho X^\dagger - \frac{1}{2} (X^\dagger X \rho + \rho  X^\dagger X) \right)$ are Lindbladian superoperators describing the relaxation of the excitonic states with a rate $\gamma$ and of the cavity modes with a total rate $\kappa_\text{tot}=\kappa+\kappa_\text{loss}$.

We have numerically solved this equation using the following parameters: $\hbar \delta_V=0$, $\hbar \delta_H=-70\, \mu eV$, $\delta_X=-\Delta_{FSS} / 2$, $\delta_Y=\Delta_{FSS} / 2$, $\hbar \Delta_{FSS}=15\, \mu eV$, $\hbar g=21\, \mu eV$, $\hbar \kappa_{tot}=120\, \mu eV$, $\kappa_{loss}=0.3\, \kappa_{tot}$, $\hbar \gamma=0.3\, \mu eV$,  $\theta=\pi/4$. The time evolution of the excitonic population in the $|H\rangle$ and $\ket{V}$ modes are computed as $P_i(t) = \langle \sigma_i^\dagger \sigma_i \rangle$ with $i=H,V$. The flip probability is
defined as the sum of the two exciton population $P_H(t)+P_V(t)$ taken at a time corresponding to its maximal value for a pulse with one photon on average. Finally the number of collected photons per pulse in the mode H is calculated as:
\begin{align}
	N_{H} = \kappa \int \langle a_H^\dag a_H \rangle dt
\end{align}

\subsection*{Excitation by a single-photon Fock state}
To account for an incoming single photon wave packet of the form $\ket{1_\xi} = \int dt \, \xi(t) \, a^\dag  \,\ket{0} $, we follow Ref.~\cite{1photon} and consider the following coupled master equations:
\begin{align} \label{fock}
	& \dot{\rho}_{11}=\mathcal{L'}[\rho_{11}]+\Omega(t)\left(\left[\rho_{01},a^\dag_V\right]-\left[\rho^\dag_{01},a_V\right] \right) \nonumber \\
	& \dot{\rho}_{01}=\mathcal{L'}[\rho_{01}]-\Omega(t)\left[\rho_{00},a_V\right] \\
	& \dot{\rho}_{00}=\mathcal{L'}[\rho_{00}] , \nonumber 
\end{align}
in which the generalized density matrices $\rho_{mn}$ are defined as
\begin{align}
	& \rho_{mn} (t) = \text{Tr}_{\text{ph}} \{U(t,t_0) \rho_s(t_0) \otimes \ket{m_\xi} \bra{n_\xi} U^\dag(t,t_0) \} ,
\end{align}
$U(t,t_0)$ being the evolution operator corresponding to the total Hamiltonian from a time $t_0$ prior to the interaction. Hence $\rho_{11}$ is the density matrix of the system when a single photon is incoming. The initial conditions are $\rho_{11} (0)= \rho_{00} (0)= \rho (0)$, $\rho_{01} (0) = 0$
and $\mathcal{L'}[\rho]$ is the Liouvillian defined by
$\rho$:
\begin{align}
	\mathcal{L'}[\rho]=-\frac{i}{\hbar}\left[\hat{H}_{s},\rho\right]+D_{\gamma,\sigma_H}[\rho] + D_{\gamma,\sigma_V}[\rho] + D_{\kappa_\text{tot},a_H}[\rho] + D_{\kappa_\text{tot},a_V}[\rho]  .
\end{align}


\begin{thebibliography}{10}
	\expandafter\ifx\csname url\endcsname\relax
	\def\url#1{\texttt{#1}}\fi
	\expandafter\ifx\csname urlprefix\endcsname\relax\def\urlprefix{URL }\fi
	\providecommand{\bibinfo}[2]{#2}
	\providecommand{\eprint}[2][]{\url{#2}}
	
	\bibitem{quantuminternet}
	\bibinfo{author}{Kimble, H.~J.}
	\newblock \bibinfo{title}{The quantum internet}.
	\newblock \emph{\bibinfo{journal}{Nature}} \textbf{\bibinfo{volume}{453}},
	\bibinfo{pages}{1023--1030} (\bibinfo{year}{2008}).
	\newblock \urlprefix\url{http://dx.doi.org/10.1038/nature07127}.
	
	\bibitem{ionphoton}
	\bibinfo{author}{StuteA.} \emph{et~al.}
	\newblock \bibinfo{title}{Quantum-state transfer from an ion to a photon}.
	\newblock \emph{\bibinfo{journal}{Nat Photon}} \textbf{\bibinfo{volume}{7}},
	\bibinfo{pages}{219--222} (\bibinfo{year}{2013}).
	\newblock \urlprefix\url{http://dx.doi.org/10.1038/nphoton.2012.358}.
	
	\bibitem{atomphoton}
	\bibinfo{author}{Birnbaum, K.~M.} \emph{et~al.}
	\newblock \bibinfo{title}{Photon blockade in an optical cavity with one trapped
		atom}.
	\newblock \emph{\bibinfo{journal}{Nature}} \textbf{\bibinfo{volume}{436}},
	\bibinfo{pages}{87--90} (\bibinfo{year}{2005}).
	\newblock \urlprefix\url{http://dx.doi.org/10.1038/nature03804}.
	
	\bibitem{NVphoton}
	\bibinfo{author}{Bernien, H.} \emph{et~al.}
	\newblock \bibinfo{title}{Heralded entanglement between solid-state qubits
		separated by three metres}.
	\newblock \emph{\bibinfo{journal}{Nature}} \textbf{\bibinfo{volume}{497}},
	\bibinfo{pages}{86--90} (\bibinfo{year}{2013}).
	\newblock \urlprefix\url{http://dx.doi.org/10.1038/nature12016}.
	
	\bibitem{QDphoton}
	\bibinfo{author}{Lodahl, P.}, \bibinfo{author}{Mahmoodian, S.} \&
	\bibinfo{author}{Stobbe, S.}
	\newblock \bibinfo{title}{Interfacing single photons and single quantum dots
		with photonic nanostructures}.
	\newblock \emph{\bibinfo{journal}{Rev. Mod. Phys.}}
	\textbf{\bibinfo{volume}{87}}, \bibinfo{pages}{347--400}
	(\bibinfo{year}{2015}).
	\newblock \urlprefix\url{http://link.aps.org/doi/10.1103/RevModPhys.87.347}.
	
	\bibitem{BookAna}
	\bibinfo{author}{Predojevi\'{c}, A.} \& \bibinfo{author}{Mitchell, M.~W.}
	\newblock \emph{\bibinfo{title}{Engineering the Atom-Photon Interaction:
			Controlling Fundamental Processes with Photons, Atoms and Solids}}.
	\newblock Nano-Optics and Nanophotonics (\bibinfo{publisher}{Springer
		International Publishing}, \bibinfo{year}{2015}).
	
	\bibitem{atacgate}
	\bibinfo{author}{Volz, T.} \emph{et~al.}
	\newblock \bibinfo{title}{Ultrafast all-optical switching by single photons}.
	\newblock \emph{\bibinfo{journal}{Nat Photon}} \textbf{\bibinfo{volume}{6}},
	\bibinfo{pages}{605--609} (\bibinfo{year}{2012}).
	\newblock \urlprefix\url{http://dx.doi.org/10.1038/nphoton.2012.181}.
	
	\bibitem{waksfewphoton}
	\bibinfo{author}{Bose, R.}, \bibinfo{author}{Sridharan, D.},
	\bibinfo{author}{Kim, H.}, \bibinfo{author}{Solomon, G.~S.} \&
	\bibinfo{author}{Waks, E.}
	\newblock \bibinfo{title}{Low-photon-number optical switching with a single
		quantum dot coupled to a photonic crystal cavity}.
	\newblock \emph{\bibinfo{journal}{Phys. Rev. Lett.}}
	\textbf{\bibinfo{volume}{108}}, \bibinfo{pages}{227402}
	(\bibinfo{year}{2012}).
	\newblock
	\urlprefix\url{http://link.aps.org/doi/10.1103/PhysRevLett.108.227402}.
	
	\bibitem{jelenafew}
	\bibinfo{author}{Englund, D.} \emph{et~al.}
	\newblock \bibinfo{title}{Ultrafast photon-photon interaction in a strongly
		coupled quantum dot-cavity system}.
	\newblock \emph{\bibinfo{journal}{Phys. Rev. Lett.}}
	\textbf{\bibinfo{volume}{108}}, \bibinfo{pages}{093604}
	(\bibinfo{year}{2012}).
	\newblock
	\urlprefix\url{http://link.aps.org/doi/10.1103/PhysRevLett.108.093604}.
	
	\bibitem{loo}
	\bibinfo{author}{Loo, V.} \emph{et~al.}
	\newblock \bibinfo{title}{Optical nonlinearity for few-photon pulses on a
		quantum dot-pillar cavity device}.
	\newblock \emph{\bibinfo{journal}{Phys. Rev. Lett.}}
	\textbf{\bibinfo{volume}{109}}, \bibinfo{pages}{166806}
	(\bibinfo{year}{2012}).
	\newblock
	\urlprefix\url{http://link.aps.org/doi/10.1103/PhysRevLett.109.166806}.
	
	\bibitem{lukin2013}
	\bibinfo{author}{Thompson, J.~D.} \emph{et~al.}
	\newblock \bibinfo{title}{Coupling a single trapped atom to a nanoscale optical
		cavity}.
	\newblock \emph{\bibinfo{journal}{Science}} \textbf{\bibinfo{volume}{340}},
	\bibinfo{pages}{1202--1205} (\bibinfo{year}{2013}).
	\newblock
	\urlprefix\url{http://www.sciencemag.org/content/340/6137/1202.abstract}.
	\newblock \eprint{http://www.sciencemag.org/content/340/6137/1202.full.pdf}.
	
	\bibitem{gaterempe}
	\bibinfo{author}{Reiserer, A.}, \bibinfo{author}{Kalb, N.},
	\bibinfo{author}{Rempe, G.} \& \bibinfo{author}{Ritter, S.}
	\newblock \bibinfo{title}{A quantum gate between a flying optical photon and a
		single trapped atom}.
	\newblock \emph{\bibinfo{journal}{Nature}} \textbf{\bibinfo{volume}{508}},
	\bibinfo{pages}{237--240} (\bibinfo{year}{2014}).
	\newblock \urlprefix\url{http://dx.doi.org/10.1038/nature13177}.
	
	\bibitem{barak}
	\bibinfo{author}{Shomroni, I.} \emph{et~al.}
	\newblock \bibinfo{title}{All-optical routing of single photons by a one-atom
		switch controlled by a single photon}.
	\newblock \emph{\bibinfo{journal}{Science}} \textbf{\bibinfo{volume}{345}},
	\bibinfo{pages}{903--906} (\bibinfo{year}{2014}).
	\newblock
	\urlprefix\url{http://www.sciencemag.org/content/345/6199/903.abstract}.
	\newblock \eprint{http://www.sciencemag.org/content/345/6199/903.full.pdf}.
	
	\bibitem{dousse}
	\bibinfo{author}{Dousse, A.} \emph{et~al.}
	\newblock \bibinfo{title}{Controlled light-matter coupling for a single quantum
		dot embedded in a pillar microcavity using far-field optical lithography}.
	\newblock \emph{\bibinfo{journal}{Phys. Rev. Lett.}}
	\textbf{\bibinfo{volume}{101}}, \bibinfo{pages}{267404}
	(\bibinfo{year}{2008}).
	\newblock
	\urlprefix\url{http://link.aps.org/doi/10.1103/PhysRevLett.101.267404}.
	
	\bibitem{1Datom}
	\bibinfo{author}{Shen, J.-T.} \& \bibinfo{author}{Fan, S.}
	\newblock \bibinfo{title}{Coherent single photon transport in a one-dimensional
		waveguide coupled with superconducting quantum bits}.
	\newblock \emph{\bibinfo{journal}{Phys. Rev. Lett.}}
	\textbf{\bibinfo{volume}{95}}, \bibinfo{pages}{213001}
	(\bibinfo{year}{2005}).
	\newblock
	\urlprefix\url{http://link.aps.org/doi/10.1103/PhysRevLett.95.213001}.
	
	\bibitem{memory}
	\bibinfo{author}{Lvovsky, A.~I.}, \bibinfo{author}{Sanders, B.~C.} \&
	\bibinfo{author}{Tittel, W.}
	\newblock \bibinfo{title}{Optical quantum memory}.
	\newblock \emph{\bibinfo{journal}{Nat Photon}} \textbf{\bibinfo{volume}{3}},
	\bibinfo{pages}{706--714} (\bibinfo{year}{2009}).
	\newblock \urlprefix\url{http://dx.doi.org/10.1038/nphoton.2009.231}.
	
	\bibitem{atommemory}
	\bibinfo{author}{Specht, H.~P.} \emph{et~al.}
	\newblock \bibinfo{title}{A single-atom quantum memory}.
	\newblock \emph{\bibinfo{journal}{Nature}} \textbf{\bibinfo{volume}{473}},
	\bibinfo{pages}{190--193} (\bibinfo{year}{2011}).
	\newblock \urlprefix\url{http://dx.doi.org/10.1038/nature09997}.
	
	\bibitem{NVcoherentcontrol}
	\bibinfo{author}{Robledo, L.}, \bibinfo{author}{Bernien, H.},
	\bibinfo{author}{van Weperen, I.} \& \bibinfo{author}{Hanson, R.}
	\newblock \bibinfo{title}{Control and coherence of the optical transition of
		single nitrogen vacancy centers in diamond}.
	\newblock \emph{\bibinfo{journal}{Phys. Rev. Lett.}}
	\textbf{\bibinfo{volume}{105}}, \bibinfo{pages}{177403}
	(\bibinfo{year}{2010}).
	\newblock
	\urlprefix\url{http://link.aps.org/doi/10.1103/PhysRevLett.105.177403}.
	
	\bibitem{50photons}
	\bibinfo{author}{Aljunid, S.~A.} \emph{et~al.}
	\newblock \bibinfo{title}{Excitation of a single atom with exponentially rising
		light pulses}.
	\newblock \emph{\bibinfo{journal}{Phys. Rev. Lett.}}
	\textbf{\bibinfo{volume}{111}}, \bibinfo{pages}{103001}
	(\bibinfo{year}{2013}).
	\newblock
	\urlprefix\url{http://link.aps.org/doi/10.1103/PhysRevLett.111.103001}.
	
	\bibitem{gazzano2013}
	\bibinfo{author}{Gazzano, O.} \emph{et~al.}
	\newblock \bibinfo{title}{Bright solid-state sources of indistinguishable
		single photons}.
	\newblock \emph{\bibinfo{journal}{Nat Commun}} \textbf{\bibinfo{volume}{4}},
	\bibinfo{pages}{1425} (\bibinfo{year}{2013}).
	\newblock \urlprefix\url{http://dx.doi.org/10.1038/ncomms2434}.
	
	\bibitem{claudon2010}
	\bibinfo{author}{Claudon, J.} \emph{et~al.}
	\newblock \bibinfo{title}{A highly efficient single-photon source based on a
		quantum dot in a photonic nanowire}.
	\newblock \emph{\bibinfo{journal}{Nat Photon}} \textbf{\bibinfo{volume}{4}},
	\bibinfo{pages}{174--177} (\bibinfo{year}{2010}).
	\newblock \urlprefix\url{http://dx.doi.org/10.1038/nphoton.2009.287}.
	
	\bibitem{lodahl1D}
	\bibinfo{author}{Arcari, M.} \emph{et~al.}
	\newblock \bibinfo{title}{Near-unity coupling efficiency of a quantum emitter
		to a photonic crystal waveguide}.
	\newblock \emph{\bibinfo{journal}{Physical Review Letters}}
	\textbf{\bibinfo{volume}{113}} (\bibinfo{year}{2014}).
	
	\bibitem{networkrempe}
	\bibinfo{author}{Ritter, S.} \emph{et~al.}
	\newblock \bibinfo{title}{An elementary quantum network of single atoms in
		optical cavities}.
	\newblock \emph{\bibinfo{journal}{Nature}} \textbf{\bibinfo{volume}{484}},
	\bibinfo{pages}{195--200} (\bibinfo{year}{2012}).
	\newblock \urlprefix\url{http://dx.doi.org/10.1038/nature11023}.
	
	\bibitem{excitonqbit}
	\bibinfo{author}{Schwartz, I.} \emph{et~al.}
	\newblock \bibinfo{title}{Deterministic writing and control of the dark exciton
		spin using single short optical pulses}.
	\newblock \emph{\bibinfo{journal}{Phys. Rev. X}} \textbf{\bibinfo{volume}{5}},
	\bibinfo{pages}{011009} (\bibinfo{year}{2015}).
	\newblock \urlprefix\url{http://link.aps.org/doi/10.1103/PhysRevX.5.011009}.
	
	\bibitem{spinqbit}
	\bibinfo{author}{Gao, W.~B.}, \bibinfo{author}{Imamoglu, A.},
	\bibinfo{author}{Bernien, H.} \& \bibinfo{author}{Hanson, R.}
	\newblock \bibinfo{title}{Coherent manipulation, measurement and entanglement
		of individual solid-state spins using optical fields}.
	\newblock \emph{\bibinfo{journal}{Nat Photon}} \textbf{\bibinfo{volume}{9}},
	\bibinfo{pages}{363--373} (\bibinfo{year}{2015}).
	\newblock \urlprefix\url{http://dx.doi.org/10.1038/nphoton.2015.58}.
	
	\bibitem{nowak}
	\bibinfo{author}{Nowak, A.~K.} \emph{et~al.}
	\newblock \bibinfo{title}{Deterministic and electrically tunable bright
		single-photon source}.
	\newblock \emph{\bibinfo{journal}{Nature Communications}}
	\textbf{\bibinfo{volume}{5}}, \bibinfo{pages}{3240} (\bibinfo{year}{2014}).
	
	\bibitem{nearoptimal}
	\bibinfo{author}{Somaschi, N.} \emph{et~al.}
	\newblock \bibinfo{title}{Near optimal single photon sources in the solid
		state}.
	\newblock \emph{\bibinfo{journal}{Arxiv 1510.06499}} .
	
	\bibitem{rabiQD}
	\bibinfo{author}{Stievater, T.~H.} \emph{et~al.}
	\newblock \bibinfo{title}{Rabi oscillations of excitons in single quantum
		dots}.
	\newblock \emph{\bibinfo{journal}{Phys. Rev. Lett.}}
	\textbf{\bibinfo{volume}{87}}, \bibinfo{pages}{133603}
	(\bibinfo{year}{2001}).
	\newblock
	\urlprefix\url{http://link.aps.org/doi/10.1103/PhysRevLett.87.133603}.
	
	\bibitem{valia}
	\bibinfo{author}{Melet, R.} \emph{et~al.}
	\newblock \bibinfo{title}{Resonant excitonic emission of a single quantum dot
		in the rabi regime}.
	\newblock \emph{\bibinfo{journal}{Phys. Rev. B}} \textbf{\bibinfo{volume}{78}},
	\bibinfo{pages}{073301} (\bibinfo{year}{2008}).
	\newblock \urlprefix\url{http://link.aps.org/doi/10.1103/PhysRevB.78.073301}.
	
	\bibitem{1photon}
	\bibinfo{author}{Gheri, K.}, \bibinfo{author}{Ellinger, K.},
	\bibinfo{author}{Pellizari, T.} \& \bibinfo{author}{Zoller, P.}
	\newblock \bibinfo{title}{Photon-wavepackets as flying quantum bits}.
	\newblock \emph{\bibinfo{journal}{Progress of Physics}}
	\textbf{\bibinfo{volume}{46}}, \bibinfo{pages}{401--415}
	(\bibinfo{year}{1998}).
	
	\bibitem{holespin}
	\bibinfo{author}{Gerardot, B.~D.} \emph{et~al.}
	\newblock \bibinfo{title}{Optical pumping of a single hole spin in a quantum
		dot}.
	\newblock \emph{\bibinfo{journal}{Nature}} \textbf{\bibinfo{volume}{451}},
	\bibinfo{pages}{441--444} (\bibinfo{year}{2008}).
	\newblock \urlprefix\url{http://dx.doi.org/10.1038/nature06472}.
	
	\bibitem{loicini}
	\bibinfo{author}{Loo, V.}, \bibinfo{author}{Lanco, L.}, \bibinfo{author}{Krebs,
		O.}, \bibinfo{author}{Senellart, P.} \& \bibinfo{author}{Voisin, P.}
	\newblock \bibinfo{title}{Single-shot initialization of electron spin in a
		quantum dot using a short optical pulse}.
	\newblock \emph{\bibinfo{journal}{Phys. Rev. B}} \textbf{\bibinfo{volume}{83}},
	\bibinfo{pages}{033301} (\bibinfo{year}{2011}).
	\newblock \urlprefix\url{http://link.aps.org/doi/10.1103/PhysRevB.83.033301}.
	
	\bibitem{rarity}
	\bibinfo{author}{Hu, C.~Y.}, \bibinfo{author}{Munro, W.~J.} \&
	\bibinfo{author}{Rarity, J.~G.}
	\newblock \bibinfo{title}{Deterministic photon entangler using a charged
		quantum dot inside a microcavity}.
	\newblock \emph{\bibinfo{journal}{Phys. Rev. B}} \textbf{\bibinfo{volume}{78}},
	\bibinfo{pages}{125318} (\bibinfo{year}{2008}).
	
\end{thebibliography}

\begin{thebibliography}{1}
	\expandafter\ifx\csname url\endcsname\relax
	\def\url#1{\texttt{#1}}\fi
	\expandafter\ifx\csname urlprefix\endcsname\relax\def\urlprefix{URL }\fi
	\providecommand{\bibinfo}[2]{#2}
	\providecommand{\eprint}[2][]{\url{#2}}
	
	\bibitem{nowak}
	\bibinfo{author}{Nowak, A.~K.} \emph{et~al.}
	\newblock \bibinfo{title}{Deterministic and electrically tunable bright
		single-photon source}.
	\newblock \emph{\bibinfo{journal}{Nature Communications}}
	\textbf{\bibinfo{volume}{5}}, \bibinfo{pages}{3240} (\bibinfo{year}{2014}).
	
	\bibitem{nearoptimal}
	\bibinfo{author}{Somaschi, N.} \emph{et~al.}
	\newblock \bibinfo{title}{Near optimal single photon sources in the solid
		state}.
	\newblock \emph{\bibinfo{journal}{Arxiv 1510.06499}} .
	
	\bibitem{loo}
	\bibinfo{author}{Loo, V.} \emph{et~al.}
	\newblock \bibinfo{title}{Optical nonlinearity for few-photon pulses on a
		quantum dot-pillar cavity device}.
	\newblock \emph{\bibinfo{journal}{Phys. Rev. Lett.}}
	\textbf{\bibinfo{volume}{109}}, \bibinfo{pages}{166806}
	(\bibinfo{year}{2012}).
	\newblock
	\urlprefix\url{http://link.aps.org/doi/10.1103/PhysRevLett.109.166806}.
	
	\bibitem{cohen-tannoudji}
	\bibinfo{author}{Cohen-Tannoudji, C.}, \bibinfo{author}{Dupont-Roc, J.} \&
	\bibinfo{author}{Grynberg, G.}
	\newblock \emph{\bibinfo{title}{Atom-Photon interactions}}
	(\bibinfo{publisher}{Wiley-VCH}).
	
	\bibitem{1photon}
	\bibinfo{author}{Gheri, K.}, \bibinfo{author}{Ellinger, K.},
	\bibinfo{author}{Pellizari, T.} \& \bibinfo{author}{Zoller, P.}
	\newblock \bibinfo{title}{Photon-wavepackets as flying quantum bits}.
	\newblock \emph{\bibinfo{journal}{Fortschritte der Physik}}
	\textbf{\bibinfo{volume}{46}}, \bibinfo{pages}{401--416}
	(\bibinfo{year}{1998}).
	
\end{thebibliography}
\end{document}